\documentclass[11pt,twoside]{article}

\usepackage{asp2006}
\usepackage{epsf}
\usepackage{psfig}
\usepackage{lscape}

\usepackage{graphics,epsfig}
\usepackage{natbib}

\begin{document}
\title{Orbital Stability of Earth-Type Planets in Binary Systems}   
\author{J. Eberle, M. Cuntz, Z. E. Musielak}   
\affil{Department of Physics, University of Texas at Arlington, Arlington, TX 76019-0059}    

\begin{abstract} 
About half of all known stellar systems with Sun-like stars consist of two or
more stars, significantly affecting the orbital stability of any planet in these
systems.  Here we study the onset of instability for an Earth-type planet that is
part of a binary system.  Our investigation makes use of previous analytical work
allowing to describe the
permissible region of planetary motion.  This allows us to establish a criterion for 
the orbital stability of planets that may be useful in the context of future
observational and theoretical studies.
\end{abstract}



\section{Introduction}

Observational evidence for the existence of planets in stellar binary
(and higher order) systems has been given by \cite{pat02}, \cite{egg04},
\cite{egg07}, and others.  \citeauthor{egg07} presented data for more
than thirty systems, mostly wide binaries, as well as several triple star
systems, with separation distances as close as 20~AU (GJ~86).
These observations are consistent with the finding that binary (and higher
order) systems occur in high frequency in the local Galactic neighborhood
\citep{duq91,lad06,rag06,bon07}. The fact
that planets in binary systems are now considered to be relatively common
is also implied by the recent detection of debris disks in various
main-sequence stellar binary systems using the {\it Spitzer Space Telescope}
\citep[e.g.,][]{tri07}.

In the last few decades, significant progress has been made in the study of
stability of planetary orbits in stellar binary systems.  Most of these
studies focused on S-type systems, where the planet is orbiting one of the
stars with the second star to be considered a perturbator.  Recently,
\cite{dav03} investigated the orbital
stability of an Earth-mass planet around a solar-mass star in the presence
of a companion star and determined the planet's ejection time for systems
with a variety of orbital eccentricities and semimajor axes.

In our previous work \citep{stu95,mus05}, we studied the stability of both
S-type and P-type orbits in stellar binary systems, and deduced orbital stability
limits for planets.  These limits were found to depend on the mass ratio between
the stellar components.  This topic was recently revisited by \cite{cun07}
and \cite{ebe07}, who used the concept of Jacobi's integral and Jacobi's constant
\citep{sze67,roy05} to deduce stringent criteria for the stability of planetary
orbits in binary systems for the special case of the ``coplanar circular restricted
three-body problem".  In this paper, we present case studies of planetary orbital
stability for different stellar mass ratios and different initial planetary distances
from its host star.

\begin{figure*}
\centering
\begin{tabular}{cc}
\epsfig{file=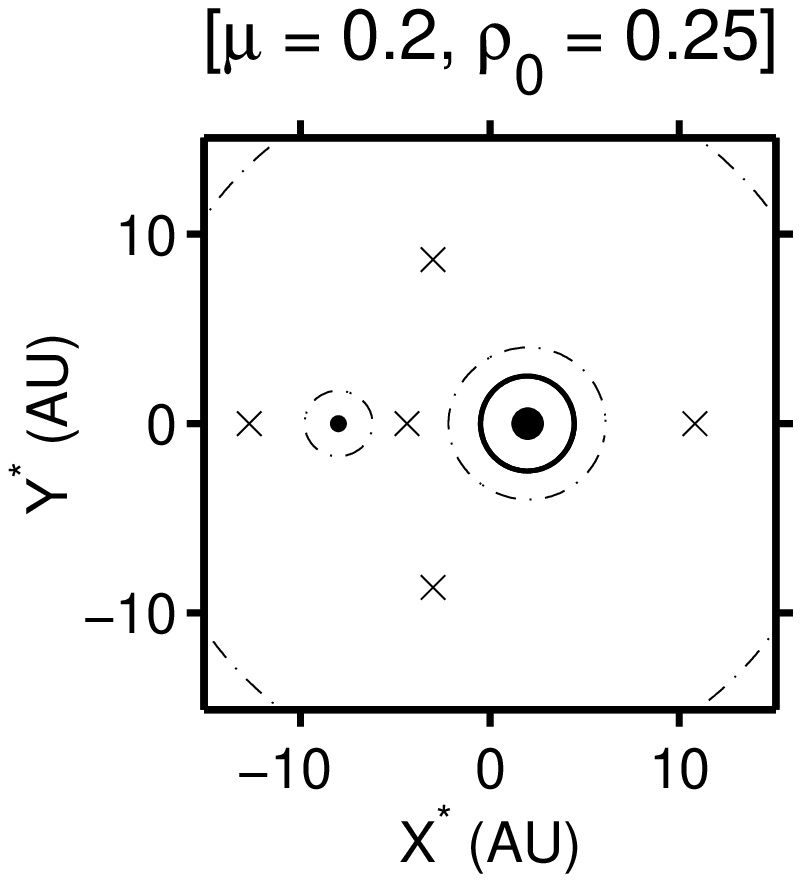,width=0.3\linewidth,height=0.335\linewidth}&
\epsfig{file=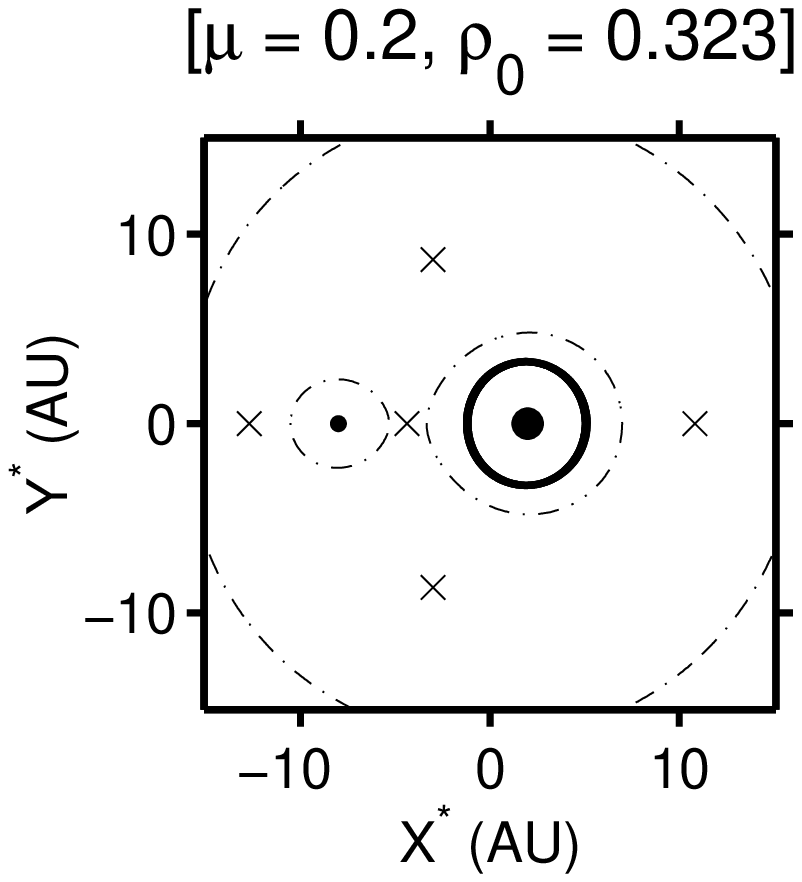,width=0.3\linewidth,height=0.335\linewidth} \\
\epsfig{file=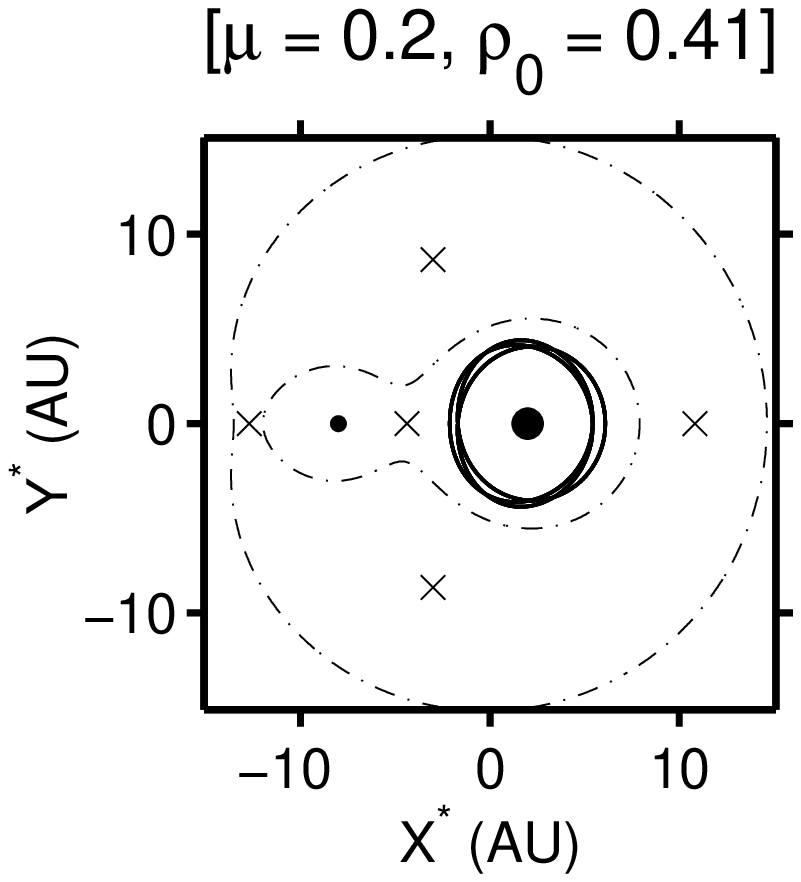,width=0.3\linewidth,height=0.335\linewidth}&
\epsfig{file=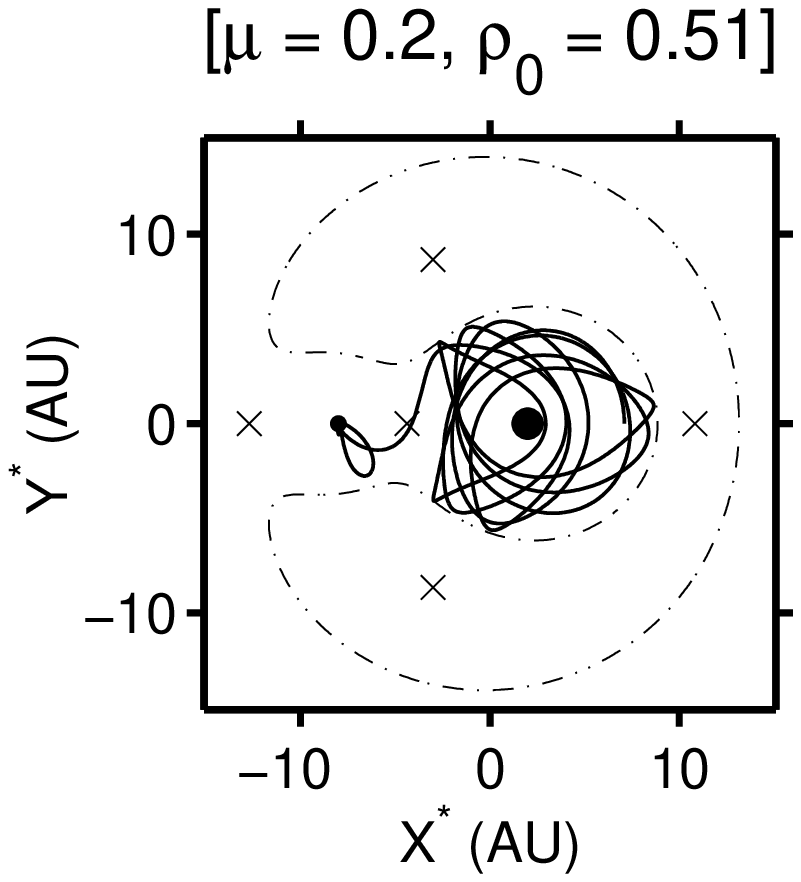,width=0.3\linewidth,height=0.335\linewidth} \\
\epsfig{file=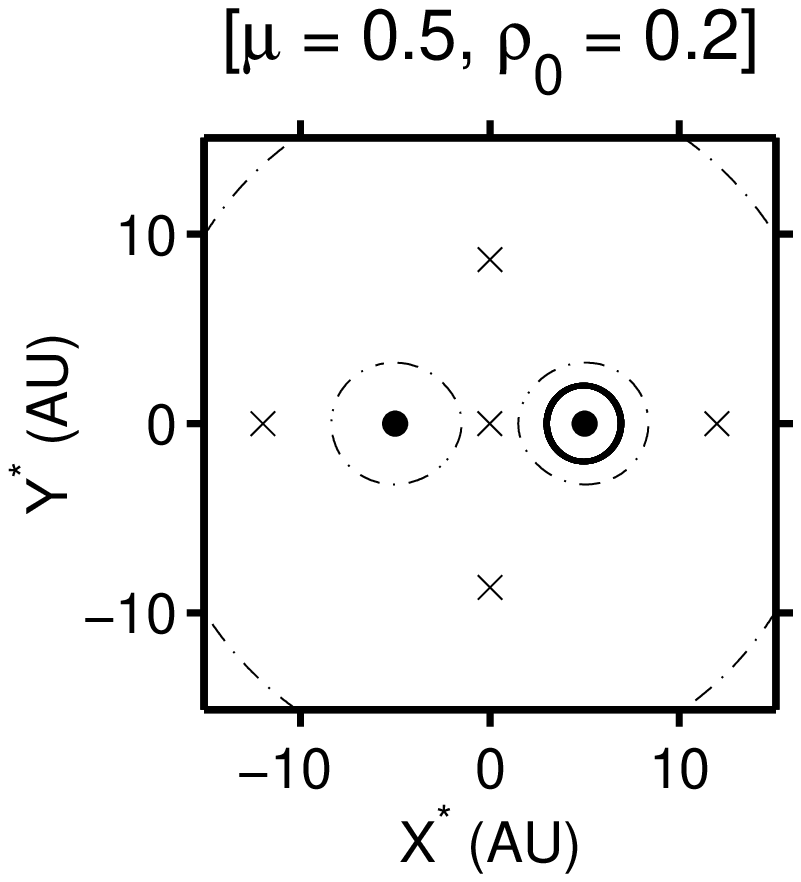,width=0.3\linewidth,height=0.335\linewidth}&
\epsfig{file=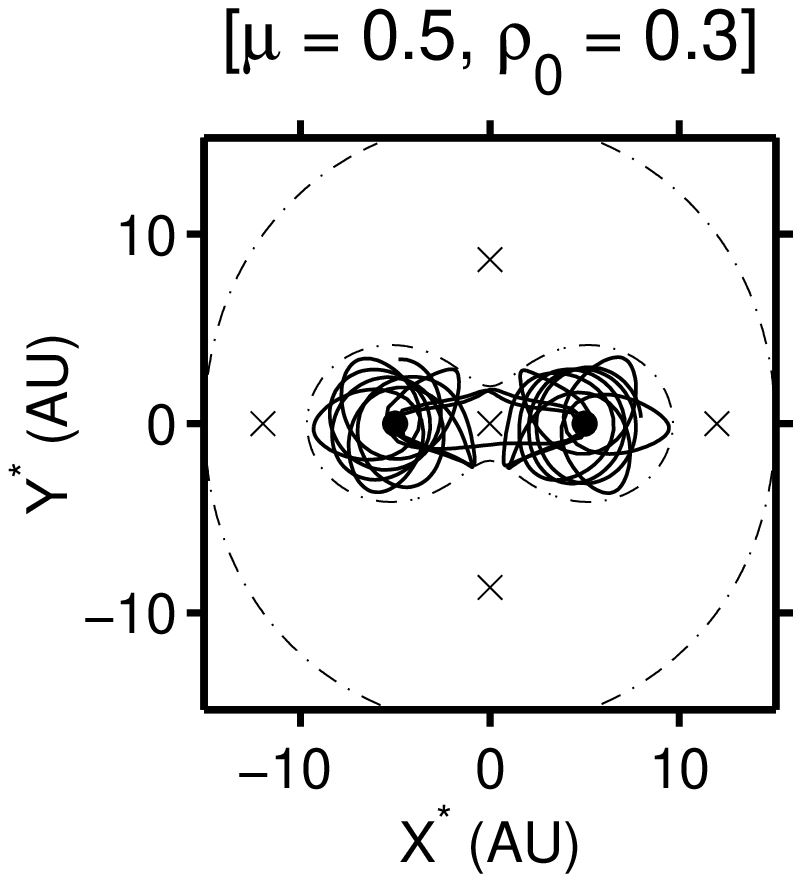,width=0.3\linewidth,height=0.335\linewidth} \\
\epsfig{file=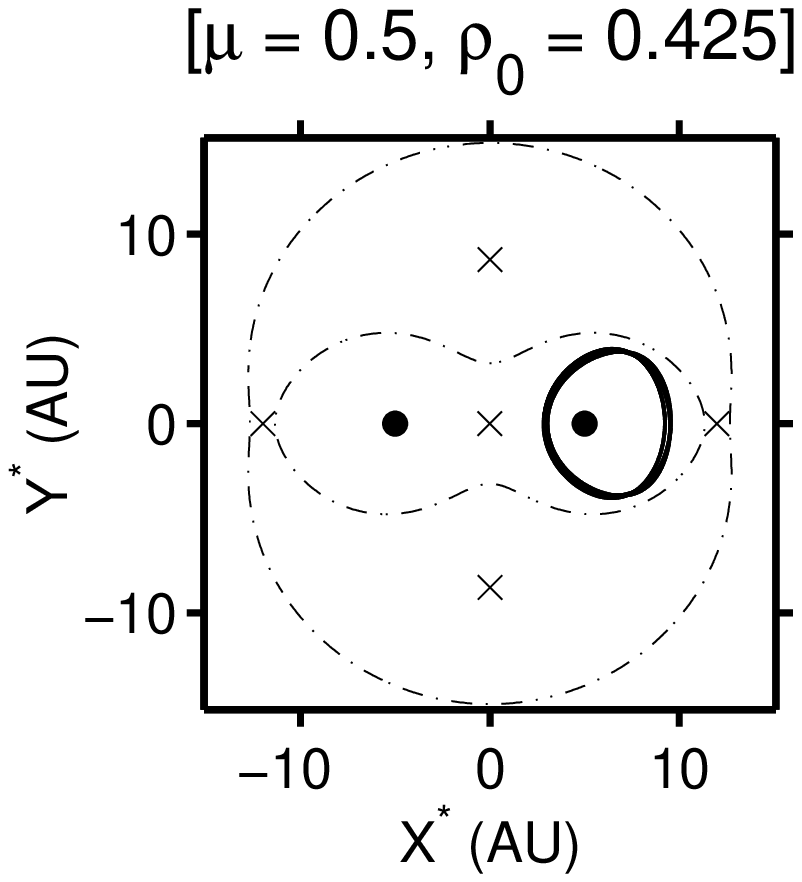,width=0.3\linewidth,height=0.335\linewidth}&
\epsfig{file=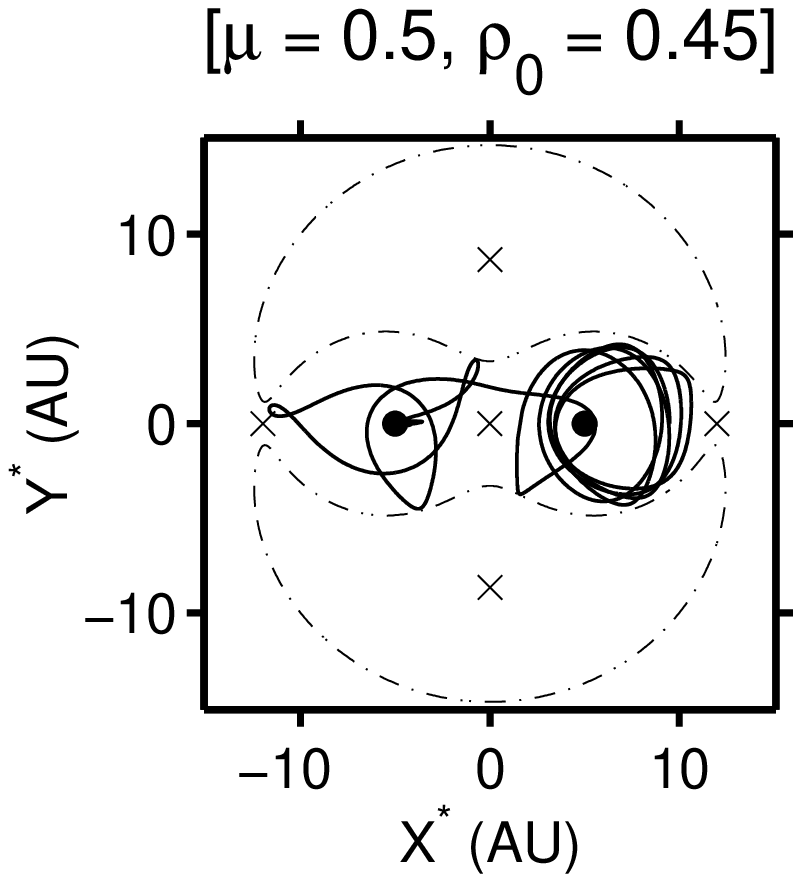,width=0.3\linewidth,height=0.335\linewidth}
\end{tabular}
\caption{Model simulations for $\mu = 0.2$ and $0.5$, and different
values of $\rho_0$.  Each panel shows the primary ({\it large dot})
and secondary ({\it small dot}) star, the planetary orbit ({\it solid line}),
the five Lagrange points, and the
``zero velocity contours" ({\it dash-dotted lines}).  The Lagrange points
are denoted as L2, L1, L3, respectively, from left to right along the line
connecting the two stars, and as L4 ({\it top}) and L5 ({\it bottom}) apart
from this line.  For $\mu = 0.2$, the critical values $\rho_0^{(1)}$,
$\rho_0^{(2)}$, and $\rho_0^{(3)}$ are given as 0.353, 0.420, and 0.692,
respectively, and for $\mu = 0.5$, they are given as 0.251, 0.442, and 0.442,
respectively.
}
\end{figure*}

\section{Methods and results}

In the so-called coplanar circular restricted three-body problem the two
stars are assumed to orbit each other in circles and their masses
are much larger than that of the planet.  In our case, it is assumed that
the planetary mass is $1 \times 10^{-6}$ of the mass of the star it orbits;
also note that the planetary motion is constrained to the orbital plane of
the two stars.  In addition, it is assumed that the initial velocity of the
planet is set for an initially circular orbit, and that it is in the same
direction as the orbital velocity of its host star.  This star shall be
the more massive of the two stars.  Furthermore, the starting position
of the planet is to the right of its host star along the line joining
the binary components (3~o'clock position).  The mass ratio $\mu$ of the
two stars is defined as $\mu = {M_2 / M}$ with $M = M_1 + M_2$, where
$M_1$ and $M_2$ are the masses of the primary and secondary star, respectively.
Additionally, $\rho_0$ denotes the planet's relative initial distance
$\rho_0 = {R_0/D}$ from its host star, with $D$ as distance between the two stars
and $R_0$ the initial planetary distance from the primary star.

In the following, we illustrate the transition from stability to instability
by progressively increasing the value of $\rho_0$ for binary systems with a
fixed mass ratio $\mu$, given as $\mu = 0.2$ and 0.5, respectively. 
For both $\mu = 0.2$ and 0.5, we present the resulting planetary orbits
for four different values of $\rho_0$ (see Fig. 1).
Each panel shows the primary (larger dot) and secondary (smaller dot)
star, the planetary orbit (solid line), as well as the zero velocity contours
(dash-dotted lines).  The upper four panels of Fig. 1 refer to the
stellar mass ratio $\mu = 0.2$, whereas the lower four panels
refer to $\mu = 0.5$, 

Let us first focus on the case studies for $\mu = 0.2$ and
$\rho_0 = 0.25$ and 0.323.  Both values of $\rho_0$
are smaller than the critical value of $\rho_0^{(1)} = 0.353$, indicating
that the planetary orbits are stable.  The fact that we restricted the
time of the simulation to 50 yrs is inconsequential owing to the fact that
the orbital stability of the planet is guaranteed by the analytical
properties of the system, namely $\rho_0 < \rho_0^{(1)}$; see \cite{cun07}
and \cite{ebe07} for a more extended discussion. 
Also note that the zero velocity contour changes between the
two panels due to the increase in $\rho_0$ by getting closer
to the Lagrange point L1, although its topology remains unaltered.
Moreover, we show two cases of unstable orbits
by choosing $\rho_0 = 0.41$ and $0.51$, respectively.  Since both
values exceed $\rho_0^{(1)}$, the zero velocity contour opens at L1,
providing the possibility for the planet to be captured by the secondary
star.  For $\rho_0 = 0.51$, the zero velocity contour even opens at L2
because of $\rho_0 > \rho_0^{(2)}$.  In case of $\rho_0 > \rho_0^{(3)}$
(not shown here), the contour would even open at L3, providing a further
type of opportunity for the planet to escape from the binary system.

The results for $\rho_0 = 0.41$ show that the
planetary orbit is unstable but still remains within the sphere of
gravitational influence of the primary star.  The situation is different
for $\rho_0 = 0.51$ where the planet reaches
the secondary star only after a few irregular orbits about the primary
star have been completed.  A more detailed analysis shows that the planet
first encountered the secondary star after 39.6 yrs at a minimal distance
of at most 0.1~AU.  A second even closer encounter occurred after 41.2 yrs,
when the simulation was stopped because the planet entered the Roche limit
of the secondary star.  The general behavior of the model with
$\rho_0 = 0.51$ is due to the fact that $\rho_0$ exceeds both $\rho_0^{(1)}$
and $\rho_0^{(2)}$, which in principle allows the planet
to escape from the binary system through the L2 point.

Our results for $\mu = 0.2$ demonstrate different cases of orbital stability
and instability.  Similar results are obtained for $\mu = 0.5$, albeit quantitative
differences due to the different values of $\mu$, $\rho_0^{(1)}$,
$\rho_0^{(2)}$, and $\rho_0^{(3)}$.  We find again that orbital
stability is obtained if $\rho_0 < \rho_0^{(1)}$, whereas for larger values of
$\rho_0$ instability is expected to emerge.  Highly unstable cases are found for
$\rho_0 = 0.3$ and 0.45.

\section{Conclusions}

For the special case of the ``coplanar circular restricted
three-body problem", we applied stringent mathematical criteria that
allow to precisely determine whether a planetary orbit in a stellar
binary system is stable or unstable.  This is accomplished by
comparing the planet's relative initial distance $\rho_0$ to the
critical values $\rho_0^{(1)}$, $\rho_0^{(2)}$,
and $\rho_0^{(3)}$, defined for a fixed stellar mass ratio $\mu$.
An adequate way of demonstrating this different type of behavior
is the assessment of the topology of the zero velocity contour, given by
the $\mu$ and $\rho_0$ values of the system.  In this case,
planetary orbital stability is obtained if the contour is completely
closed around the primary star and planet.  All numerical case studies
show a behavior consistent with this theoretical prediction. 
Note however that for generalized binary systems, other methods are
required to determine the long-term stability of
planetary orbits \citep[e.g.,][]{hol99,dav03}.  Important applications
of our work include contesting numerically deduced stability limits
for cases where analytically deduced results exist.

\acknowledgements 
This work has been supported by the Alexander von Humboldt Foundation
(Z.~E.~M.).



\begin{thebibliography}{}

\bibitem[Bonavita \& Desidera(2007)]{bon07}
Bonavita, M. \& Desidera, S. 2007, \aap, 468, 721

\bibitem[Cuntz et al.(2007)]{cun07}
Cuntz, M., Eberle, J., \& Musielak, Z. E. 2007, \apjl, 669, L105

\bibitem[Danby(1988)]{dan88}
Danby, J. M. A. 1988, Fundamentals of Celestial Mechanics
(Richmond: Willmann-Bell, Inc.)

\bibitem[David et al.(2003)]{dav03}
David, E.-M., Quintana, E. V., Fatuzzo, M., \& Adams, F. C. 2003, \pasp, 115, 825

\bibitem[Duquennoy \& Mayor(1991)]{duq91}
Duquennoy, A., \& Mayor, M. 1991, \aap, 248, 485

\bibitem[Eberle et al.(2007)]{ebe07}
Eberle, J., Cuntz, M., \& Musielak, Z. E. 2007, \aap, submitted

\bibitem[Eggenberger \& Udry(2007)]{egg07}
Eggenberger, A., \& Udry, S. 2007, in Planets in Binary Star Systems,
ed. Haghighipour (New York: Springer), in press

\bibitem[Eggenberger et al.(2004)]{egg04}
Eggenberger, A., Udry, S., \& Mayor, M. 2004, \aap, 417, 353

\bibitem[Holman \& Wiegert(1999)]{hol99}
Holman, M. J., \& Wiegert, P. A. 1999, \aj, 117, 621

\bibitem[Lada(2006)]{lad06}
Lada, C. J. 2006, \apjl, 640, L63

\bibitem[Musielak et al.(2005)]{mus05}
Musielak, Z. E., Cuntz, M., Marshall, E. A., \& Stuit, T. D. 2005,
\aap, 434, 355

\bibitem[Patience et al.(2002)]{pat02}
Patience, J., et al. 2002, \apj, 581, 654

\bibitem[Quintana et al.(2002)]{qui02}
Quintana, E. V., Lissauer, J. J., Chambers, J. E., \& Duncan, M. J. 2002,
\apj, 576, 982

\bibitem[Raghavan et al.(2006)]{rag06}
Raghavan, D., et al. 2006, \apj, 646, 523

\bibitem[Roy(2005)]{roy05}
Roy, A. E. 2005, Orbital Motion (Bristol and Philadelphia: Institute of
Physics Publ.)

\bibitem[Stuit(1995)]{stu95}
Stuit, T. D. 1995, M.S. thesis, University of Alabama in Huntsville

\bibitem[Szebehely(1967)]{sze67}
Szebehely, V. 1967, Theory of Orbits (New York and London: Academic
Press)

\bibitem[Trilling et al.(2007)]{tri07}
Trilling, D. E., et al. 2007, \apj, 658, 1289

\end{thebibliography}
\end{document}